\documentclass[superscriptaddress,twocolumn]{revtex4}
\usepackage{amssymb}
\usepackage[tbtags]{amsmath}
\usepackage{graphicx}
\usepackage{epsfig,graphicx,times}

\newcommand{\ket}[1]{|#1\rangle}

\begin{document}

\title{Testing the Kochen-Specker theorem with Josephson qubits}
\author{L. F. Wei}
\affiliation{Laboratory of Quantum Opt-electronic Information,
Southwest Jiaotong University, Chengdu 610031, China}

\affiliation{Advanced Science Institute, The Institute of Physical
and Chemical Research (RIKEN), Wako-shi, Saitama, 351-0198, Japan}

\affiliation{CREST, Japan Science and Technology Agency (JST),
Kawaguchi, Saitama 332-0012, Japan}

\author{K. Maruyama} \affiliation{Advanced Science Institute, The Institute of
Physical and Chemical Research (RIKEN), Wako-shi, Saitama, 351-0198, Japan}
\author{X.-B. Wang}
\affiliation{Advanced Science Institute, The Institute of Physical
and Chemical Research (RIKEN), Wako-shi, Saitama, 351-0198,
Japan}

\affiliation{Department of Physics, Tsinghua University, Beijing
100084, China}

\author{J. Q. You}
\affiliation{Advanced Science Institute, The Institute of Physical
and Chemical Research (RIKEN), Wako-shi, Saitama, 351-0198,
Japan}

\affiliation{Department of Physics and Surface Physics Laboratory
(National Key Laboratory), Fudan University, Shanghai 200433, China}

\author{Franco Nori}
\affiliation{Advanced Science Institute, The Institute of Physical and Chemical Research (RIKEN),
Wako-shi, Saitama, 351-0198, Japan}
\affiliation{Center for Theoretical Physics, Physics Department, CSCS, The University of Michigan,
Ann Arbor, Michigan 48109-1040, USA}
\date{\today}

\begin{abstract}
We propose an experimental approach to {\it macro}scopically test
the Kochen-Specker theorem (KST) with superconducting qubits. This
theorem, which has been experimentally tested with single photons
or neutrons, concerns the conflict between the contextuality of
quantum mechnaics (QM) and the noncontextuality of hidden-variable
theories (HVTs). We first show that two Josephson charge qubits
can be controllably coupled by using a two-level data bus produced
by a Josephson phase qubit.
Next, by introducing an approach to perform the expected joint
quantum measurements of two separated Josephson qubits, we show that
the proposed quantum circuits could demonstrate quantum
contextuality by testing the KST at a macroscopic level.

PACS number(s): 03.65.Ta, 
03.67.Lx, 
85.25.Dq. 

\end{abstract}

\maketitle

{\it Introduction.---}Quantum measurements are of a statistical
nature. There have been proposals for a more complete description of
quantum systems in terms of so-called hidden-variable theories
(HVTs)~\cite{book}, where all observables have definite values at
all times. Historically, two important theorems (among many others),
have been proposed by Bell and Kochen-Specker to elucidate the
incompatibility between the predictions of quantum mechnaics (QM)
and those of HVTs~\cite{book,rev1}. Bell's theorem states that,
given a premise of locality, a HVT cannot match the statistical
predictions of QM, while the Kochen-Specker theorem
(KST)~\cite{book,rev1,ks1} claims that the contextuality predicated
by QM (i.e., the measured result of an observable dependes on the
experimental context, in which other co-measurable observables are
measured simultaneously) conflicts with the noncontextuality in HVTs
(wherein the results of measurements are independent of the order
the measurements are performed).

Many experiments~\cite{Bell} have demonstrated the existence of
various non-local correlations that cannot be explained with
reference to any ``local" theory in classical physics. However, to
our knowledge, only two kinds of experiments~\cite{kse} with
single photons and single neutrons, respectively, have already
been demonstrated to test the KST. This is because that the
experimental feasible test of the KST with two qubits requires a
more stringent implementation~\cite{book,ks1,kse}, i.e., {\it
joint} measurements (instead of just the {\it independent} ones
for testing Bell's theorem) on the qubits to obtain the results of
commuting observables. The aim of this work is to provide a
possible way to test the KST at a macroscopic level by using
superconducting quantum circuits~\cite{coupling-1} and appropriate
joint quantum measurements on two macroscopic qubits.

The two qubits used in previous tests~\cite{kse} of the KST were
encoded by two degrees of freedom (i.e., the path- and
polarization components) of {\it single} photons or neutrons. The
present qubits, i.e., Josephson charge qubits (JCQs), are
generated by {\it two macroscopic} ``particles"---Cooper-pair
boxes (CPBs) with about $10^{9}$ Cooper pairs~\cite{nec1}.
Desirable controllable inter-qubit couplings could be implemented
by coupling the qubits to a common data bus, a two-level system
produced also by another macroscopic qubit: a Josephson phase
qubit (JPQ)~\cite{martinis1}. Indirectly coupling JCQs (rather
than directly coupling them either capacitively or inductively)
provides an obvious advantage to perform desirable independent
measurements on the two qubits. In most indirect-coupling schemes,
the inter-qubit interactions are usually mediated by {\it bosonic}
modes, e.g., cavity modes for atomic qubits, the center-of-mass
vibrational modes for trapped ions, or LC-oscillator modes for
Josephson qubits~\cite{bs1, bs2}. Here we propose an alternative
approach to indirectly couple JCQs by utilizing a different type
of data bus, hereafter called a two-level data bus (TLDB),
produced by a {\it two-level} system such as a JPQ. Recently, the
controllable coupling between a JCQ~\cite{nec1} and a
JPQ~\cite{martinis1} has been experimentally
demonstrated~\cite{cp-coup}. Thus, coupling two JCQs by a JPQ
should be experimentally feasible.

The joint measurements of two qubits (using the path- and
polarization components of a single photon or neutron) in the
previous tests of the KST~\cite{kse} were demonstrated by {\it
successively} using a sequence of filters, e.g., polarizing
beamsplitters for a photon and spin analyzers for a neutron. Here,
desirable joint measurements would be achieved by combining two
independent measurements performed {\it simultaneously} on two
uncoupled and {\it not-moving} CPBs, rather than the {\it
fast-escaping} photons or neutrons. For example, an
$X_1$-measurement ($\sigma_1^x$) and an $Z_2$-measurement
($\sigma_2^z$) could be combined as a joint measurement
$J_1\,(=Z_1X_2)$ by using {\it just a single detector}~\cite{wang}.
By introducing a measured circuit with two dc superconducting
quantum interference device (dc-SQUID) detectors, joint measurements
of two commuting observables (such as $J_1$ and $J_2=Z_2X_1$) could
be simultaneously implemented. As a consequence, the KST should be
tested with the proposed macroscopic superconducting quantum
circuits.

{\it Controllable coupling between JCQs.---}We consider the
quantum circuit shown in Fig.~1, wherein two SQUID-based CPBs are
connected to a common bus, i.e., a current-biased Josephson
junction (CBJJ). The $k$th ($k=1,2$) CPB is biased by an external
flux $\Phi_k$ and a gate voltage $V_k$, and the CBJJ is biased by
a dc current $I_b$. We assume that the two CPBs have equal
junction capacitances (i.e., $c_{J1}=c_{J2}$), gate capacitances
($C_{g1}=C_{g2}$), and also are biased by the same external
voltages: $V_1=V_2$. Therefore, there is no {\it direct} coupling
between these two CPBs, but there is an {\it indirect} interaction
via the CBJJ. The coupling between the $k$th CPB and the CBJJ
results from the voltage relation: $V_{k}=V_{Jk}+V_b+V_{gk}$, with
$V_{Jk},\,V_{gk}$,\, and $V_{b}$ being the voltages across the
junctions, the gate capacitance of the $k$th CPB, and the CBJJ,
respectively. This circuit can be easily generalized to include
more qubits, coupled by a common CBJJ.
The Hamiltonian of this circuit is~\cite{bs1}
\begin{equation}\label{ham_total}
\hat{H}=\hat{H}_1+\hat{H}_2+\hat{H}_b+\hat{H}_{1b}+\hat{H}_{2b},
\end{equation}
where
$\hat{H}_k=2e^2(\hat{n}_k-n_{gk})^2/C_k-E_{Jk}(\Phi_k)\cos\hat{\theta}_k,\,k=1,2$,\,\,
$\hat{H}_{b}=\hat{p}_b^2/[2\tilde{C}_b(\Phi_0/2\pi)^2]-E_{jb}(\cos\hat{\theta}_b-I_b\hat{\theta}_k/I_0)$,\,
and $\hat{H}_{kb}=2\pi
C_{gk}E_{Jk}(\Phi_k)\hat{\theta}_b\sin\hat{\theta}_k/(C_k\Phi_0)$
are the effective Hamiltonians describing the $k$th CPB, the CBJJ,
and the coupling between them, respectively.
$E_{Jk}(\Phi_k)=2\epsilon_{Jk}\cos(2\pi\Phi_k/\Phi_0)$ and
$C_k=2c_{Jk}+C_{gk}$ are the effective Josephson energy and
capacitance of the $k$th CPB. Also, $E_{Jb}$ and
$\tilde{C}_b=C_{Jb}+\sum_{k=1}^2(C_{Jk}^{-1}+C_{gk}^{-1})^{-1}$ are
the Josephson energy and effective capacitance of the CBJJ,
respectively. The operators $\hat{n}_k$ and $\hat{\theta}_k$,
satisfying the commutation relations $[\hat{\theta}_k,\hat{n}_k]=i$,
describe the excess number of Cooper pairs and the effective phase
across the junctions in the $k$th CPB, respectively. In addition,
the phase operator $\hat{\theta}_b$ for the CBJJ and its conjugate
$\hat{p}_b$ satisfy another commutation relation
$[\hat{\theta}_b,\hat{p}_b]=i\hbar$. Finally, we note that the
coupling between the $k$th CPB and the CBJJ is controllable; it can
be switched on/off by just switching on/off the effective Josephson
energy of the $k$th CPB, via adjusting the external flux $\Phi_k$
applied to the $k$th SQUID-loop.

\begin{figure}[tbp]
\vspace{0.5cm}
\includegraphics[width=13.8cm, height=6.8cm]{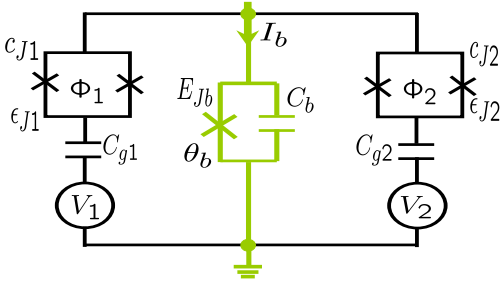}
\vspace{-3.8cm}
\caption{(Color online) Two
Josephson charge qubits (JCQs) are controllably coupled to a common
current-biased Josephson junction (CBJJ) (denoted by the dark green
part), which operates as a Josephson phase qubit (JPQ) and acts as a
coupler.}
\end{figure}

Suppose that the CPBs are biased such that
$n_{gk}=C_{gk}V_k/(2e)\sim 1/2$, and thus they behave as effective
two-level systems (with the basis
$\{\ket{0_k},\ket{1_k}\},\,k=1,2$) generating JCQs. By introducing
Pauli operators defined in terms of this excess-charge-state
basis, the $k$th JCQ has the Hamiltonian
$\hat{H}_k=4e^2(n_{gk}-1/2)\hat{\sigma}_k^z/2C_k-E_{Jk}(\Phi_k)\hat{\sigma}_k^x/2$.
On the other hand, it is well known that a CBJJ can be approximated
as a harmonic oscillator~\cite{bs1}, if it is biased as $I_b\ll
I_0=2\pi E_{Jb}/\Phi_0$. Here, we consider a different case, where
the biased dc current $I_b$ is slightly smaller than the critical
current $I_0$, and thus the CBJJ has only a few bound states. The
two lowest energy states, $|0_b\rangle$ and $|1_b\rangle$, are
selected to define a JPQ acting as a TLDB. Under such condition, the
Hamiltonian of the CBJJ reduces to
$\hat{H}_b=\hbar\omega_b\hat{\sigma}_b^z$, with
$\hat{\sigma}_b^z=|0_b\rangle\langle 0_b|-|1_b\rangle\langle 1_b|$
being the standard Pauli operator and $\omega_b$ the eigenfrequency.

The controllability of the present quantum circuit is due to the
fact that the external flux and voltage biases for the JCQs are
manipulable. For example, the charging energy
$E_k^C(n_{gk})=4e^2(n_{gk}-1/2)/C_k$ of the $k$th JCQ can be
switched off by setting the gate voltage $V_k$ such that
$n_{gk}=1/2$. Also, by adjusting the external flux $\Phi_k$ one can
control the effective Josephson energy of the $k$th qubit and
consequently its coupling to the JPQ.
By setting $n_{g1}=n_{g2}=1/2$ and
$E_{J1}(\Phi_1)\omega_b,\,E_{J1}(\Phi_1)\omega_b>0,\,$ the above
Hamiltonian (1) reduces to (under the usual rotating wave
approximation in the interaction picture)
\begin{eqnarray}\label{ham_rwa}
\tilde{H}_1(t)=\sum_{k=1}^2\left\{\lambda_k(\Phi_k)\tilde{\sigma}_k^\dagger\hat{\sigma}_b^-\exp[-i\Delta_k(\Phi_k)t]+H.c.\right\},
\end{eqnarray}
where $\lambda_k(\Phi_k)=2i\pi\theta_b^{01}
C_{gk}E_{Jk}(\Phi_k)/(C_k\Phi_0)$, and
$\Delta_k(\Phi_k)=\omega_b-E_{Jk}(\Phi_k)/\hbar$ are the coupling
strength and the detuning between the $k$th JCQ and the JPQ,
respectively. $\theta_b^{kj}=\langle
k_b|\hat{\theta}_b|j_b\rangle\,(k,j=0,1)$\, are the
``electric-dipole" matrix elements for the TLDB. The ladder
operators in Eq.~(\ref{ham_rwa}) are defined by
$\tilde{\sigma}_k^\dagger=|+_k\rangle\langle
-_k|$,\,$|\pm_k\rangle=(|0_k\rangle\pm|1_k\rangle)/\sqrt{2}$, and
$\hat{\sigma}_b^\dagger=|0_b\rangle\langle 1_b|$.
The JPQ can serve as a TLDB to transport information between the two
JCQs. By switching on the Josephson energy of one of the JCQs (just
varying the applied $\Phi_k$), the JCQ can be tunably coupled to the
TLDB with {\it fixed} parameters. As a consequence, for example,
quantum information stored in these two JCQs can be exchanged by
sequentially performing two SWAP gates; one between the TLDB and the
$k$th JCQ and then another between the TLDB and the $j$th one
($k\neq j$).

The indirect coupling between the JCQs could also be designed to
produce a direct {\it dynamical} interaction between them
(although there is no direct coupling between them) by
adiabatically eliminating the commonly-connected TLDB. This has
been widely done with bosonic data buses before~\cite{blais2}, but
never with TLDB. Indeed, by controlling the Josephson energies of
the qubits (such that $E_{J1}(\Phi_1)\omega_b>0$, with
$E_{J2}(\Phi_1)\omega_b<0$), the interaction Hamiltonian
(\ref{ham_rwa}) can be replaced by
\begin{eqnarray}\label{ham03}
\tilde{H}_2(t)&=&\lambda_1(\Phi_1)\tilde{\sigma}_1^\dagger\hat{\sigma}_b^-\exp[-i\Delta_1(\Phi_1)t]\nonumber\\
&+&\lambda_2(\Phi_2)\tilde{\sigma}_2^\dagger\hat{\sigma}_b^+\exp[-i\Delta_2(\Phi_2)t]+H.c.
\end{eqnarray}
We further assume that the external fluxes are properly set so that
$|E_{J1}(\Phi_1)|=|E_{J2}(\Phi_2)|=E_J$, yielding
$|\lambda_1(\Phi_1)|=|\lambda_2(\Phi_2)|=\lambda,\,|\Delta_1(\Phi_1)|=|\Delta_2(\Phi_2)|=\Delta$.
Here, we consider the large-detuning regime, with $\lambda/\Delta\ll
1$, which can be easily satisfied for the typical experimental
parameters (e.g.~\cite{martinis1}, $\lambda$ is usually less than a
few hundred MHz, while $\Delta$ could be adjusted to a few GHZ).
Thus, the Hamiltonian (\ref{ham03}) can be approximated as
$\tilde{H}_{3}=\lambda^2\hat{\sigma}_b^z(\tilde{\sigma}_1^\dagger\tilde{\sigma}_2^\dagger+H.c.)/\Delta$.
This implies that the coupler (i.e., the TLDB simultaneously
connecting to the two qubits) could be adiabatically eliminated,
since its excitation is virtual. For example, if the TLDB is
initially prepared in its ground state, then it always remains in
its ground state. Therefore, the above three-body Hamiltonian
$\tilde{H}_{3}$ can be effectively reduced to
\begin{equation}\label{ham_eff}
\hat{H}_{\rm
dyn}=\frac{\lambda^2}{\Delta}(\tilde{\sigma}_1^\dagger\tilde{\sigma}_2^\dagger+H.c.),
\end{equation}
which describes a dynamically-induced direct interaction between the
above two JCQs.

In this circuit, single-qubit operations are relatively simple. For
example, a $\sigma^x_k$-rotation
$\tilde{R}_k^x(\beta)=\exp(i\beta\tilde{\sigma}_k^x)$, where
$\beta=2e^2(n_{gk}-1/2)t/(\hbar C_k)$, can be implemented by making
its Josephson energy vanish (thus decoupling the qubit from the
TLDB) and varying the gate voltage $V_k$ slightly from its
degeneracy point (i.e., $V_k=e/C_{gk}$). Meanwhile, if only the
$k$th qubit is coupled to the TLDB in the large detuning regime (via
adjusting the applied fluxes), with
$|\lambda_k(\Phi_k)/\Delta_k(\Phi_k)|\ll 1$, then a
$\sigma^z_k$-rotation can be implemented as
$\tilde{R}_k^z(\gamma)=\exp(i\gamma\tilde{\sigma}_k^z)$, with
$\gamma=|\lambda_k(\Phi_k)|^2t/[2\hbar|\Delta_k(\Phi_k)|]$.

{\it Joint measurements for testing the KST.---}Following the logic
used in~\cite{simon} the test of the KST requires a composite
quantum system (consisting of subsystems 1 and 2) or a single system
with two degrees of freedom for which $(i)$ one always finds the
same outcomes for two sets of co-measurable (i.e., simultaneously
measurable) dichotomic (e.g., $\pm 1$) observables $\{Z_1,\,Z_2\}$
(i.e., $v(Z_1)=v(Z_2)$) and $\{X_1,\,X_2\}$ (i.e., $v(X_1)=v(X_2)$),
and $(ii)$ one can perform joint measurements $J_1=Z_1X_2$ and
$J_2=X_1Z_2$ that are co-measurable as well.

Noncontextuality in HVTs implies that all the observables of a
system have predetermined noncontextual values. This means that the
value $v_1$ (or $v_2$) of the joint measurement $J_1$ (or $J_2$) is
given as the product of the values of each independent observables,
namely, $v_1=v(Z_1)v(X_2)$ (or $v_2=v(Z_2)v(X_1)$). Also, the value
of the joint measurement should be independent of the experimental
context, i.e.,
\begin{equation}
{\rm HVT}: \,\,\,v_1\,v_2\,=\,1.
\end{equation}
On the other hand, in QM there exists a quantum state
$|\psi_{12}\rangle$ that gives the same outcomes for the observables
$\{Z_1,Z_2\}$ and also for $\{X_1,X_2\}$. This state is also an
eigenstate of $J_1J_2$ with the eigenvalue $-1$, i.e.,
$(X_1Z_2)(X_2Z_1)|\psi_{12}\rangle=-|\psi_{12}\rangle$.
Thus, the measured value $v_1$ of the observable $J_1$ on this state
will always have opposite sign to that $v_2$ of $J_2$, i.e.,
\begin{equation}
{\rm QM}:\,\,\, v_1\,v_2\,=\,-1.
\end{equation}
Therefore, the noncontextuality in a HVT [Eq.~(5)] is incompatible
with the contextuality in standard QM [Eq.~(6)]. A generic proposal
to test such a conflict is shown in Fig.~2(a), wherein $v_1$ and
$v_2$ readout the joint measurements $J_1$ and $J_2$. By combining
[denoted by the red dotted-line part in Fig.2(a)] the values of
$v_1$ and $v_2$, we can check either $v_1\,v_2=1$ or $v_1v_2=-1$ to
implement the test.

Our proposal for testing the KST (with the macroscopic circuit
proposed above) consists of the following three steps:

(1) Prepare a quantum state of a composite system for which the measured results of $Z_1$ and
$Z_2$ are always found to be equal to each other, and the same for $X_1$ and $X_2$.

The effective Hamiltonian in Eq.~(\ref{ham_eff}) can directly
deliver such a quantum state, and the dichotomic observables can be
defined as:
$X_k=\tilde{\sigma}_k^x=\hat{\sigma}_k^z,\,Z_k=\tilde{\sigma}_k^z=\hat{\sigma}_k^x\,(k=1,2)$.
The time evolution operator for Eq.~(\ref{ham_eff}) can then be
expressed as:
%
$\tilde{U}_{\rm dyn}(\alpha)=\cos\alpha(|\!--\rangle\langle
--\!|+|\!++\rangle\langle ++\!|)+i\sin\alpha(|\!--\rangle\langle
++\!|-|\!++\rangle\langle --\!|)$,\,with
$\alpha=\lambda^2t/\hbar\Delta$.
Thus, starting with the initial state
$|\psi(0)\rangle=|\!--\rangle$, the above two-qubit evolution,
followed by a $\sigma_k^z$-rotation, can generate the desired
entangled state,
\begin{equation}
|\psi_{12}\rangle=\tilde{R}_1^z\left(\frac{\pi}{4}\right)\tilde{U}_{\rm
dyn}\left(\frac{3\pi}{4}\right)|\!--\rangle=\frac{1}{\sqrt{2}}(|\!--\rangle+|\!++\rangle).
\end{equation}

(2) Perform the joint measurement of $Z_1$ and $Z_2$, and also of
$X_1$ and $X_2$, to confirm the above requirement {\it $(i)$}, i.e.,
\begin{equation}
Z_1Z_2|\psi_{12}\rangle=X_1X_2|\psi_{12}\rangle=|\psi_{12}\rangle.
\end{equation}
For the quantum circuit proposed above the measurements of $Z_k$ and
$X_k$ could be experimentally performed by individually detecting
the circulating current $I_k^s$ (i.e., $\hat{I}_k^s\simeq
I_c\hat{\sigma}_k^x=I_c\tilde{\sigma}_k^z,\,I_c=2\pi\epsilon_J/\Phi_0$)
along the $k$th SQUID-loop (when it decouples from the TLDB by
setting $\Phi_k=\Phi_0/2$) and the excess charge $n_k$ (i.e.,
$\tilde{\sigma}_k^x=\hat{\sigma}_k^z=|0_k\rangle\langle
0_k|-|1_k\rangle\langle 1_k|$) on the $k$th CPB, respectively.
Although the present qubits work in the charge regime, the above
critical current $I_c$ could still reach an experimentally
measurable value, e.g., $\sim 8$ nA for a typical Josephson
junction~\cite{nec1} with $\epsilon_J\sim 25\,\mu$eV.

(3) Design an experimentally feasible approach to simultaneously
perform two joint measurements of $Z_1X_2$ and $Z_2X_1$ for testing
the conflict between the contextuality in QM and the
noncontextuality in HVTs.

The quantum state prepared above could be rewritten as~\cite{simon}
$|\psi_{12}\rangle=(|\chi_{1,-1}\rangle+|\chi_{-1,1}\rangle)/\sqrt{2}$,
with $|\chi_{1,-1}\rangle$ and $|\chi_{-1,1}\rangle$ being two
normalized eigenstates of the commuting joint operators $J_1$ and
$J_2$. This implies that, when we perform the above joint
measurements, the state of the quantum circuit will collapse to
either $|\chi_{1,-1}\rangle$ or $|\chi_{1,-1}\rangle$. The first
(second) index of $\xi$ indicates the eigenvalue with respect to
$J_1$\,($J_2$).
The state
$|\chi_{1,-1}\rangle=(|00\rangle+|11\rangle+|10\rangle-|01\rangle)$
implies that if the simultaneous measurements of $X_1$ and $X_2$
show the same results (i.e., $X_1X_2=1$), then continuously
performing the simultaneous measurements $Z_2$ and $Z_1$ always
induce opposite results: one is $|+\rangle$ and another must be
$|-\rangle$. Similar arguments can also be obtained for the state
$|\chi_{-1,1}\rangle=(|00\rangle+|11\rangle-|10\rangle+|01\rangle)$.
Therefore, regardless of if the system collapses to either the state
$|\chi_{1,-1}\rangle$ or $|\chi_{1,-1}\rangle$, the results of its
two joint measurements, $v_1=v(Z_1X_2)$ and $v_2=v(Z_2X_1)$, are
always opposite.
This is a clear contradiction with Eq.~(5), which is predicted by
HVTs.

A specific approach to test the KST by simultaneously performing two
joint measurements $Z_1X_2$ and $Z_2X_1$ on two JCQs is shown in
Fig.~2(b). Here, the coupling between the two JCQs (with the same
gate voltage biases $V_1=V_2=V$) is switched off for individual
detections, by setting $\Phi_k=\Phi_0/2,\,k=1,2$.
\begin{figure}[tbp]
\vspace{0.5cm}
\includegraphics[width=12.8cm, height=6.8cm]{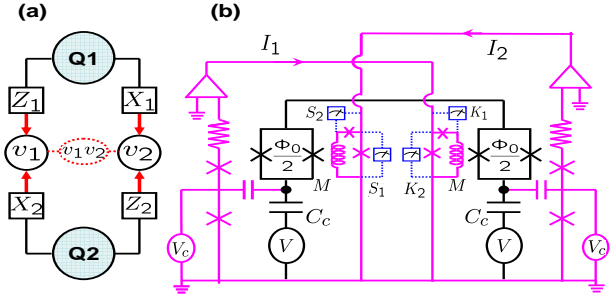}
\vspace{-3.5cm}
\caption{(Color online) (a) A schematic diagram of
joint measurements on two qubits ($Q1$ and $Q2$): $Z_i$ and $X_i$
($i=1,2$) refer to the $Z$- and $X$-measurements on the $i$th qubit
$Q_i$, respectively; $v_1$ and $v_2$ detect the joint observables
$J_1=Z_1X_2$ and $J_2=X_1Z_2$, respectively. The KST is tested by
comparing (red dotten-line part) these detections to check either
$v_1\,v_2=1$ (predicted by HVT) or $v_1\,v_2=-1$ (predicted by QM).
(b) A specific approach to implement desirable joint measurements to
test the KST with two JCQs. The colored parts refer to the proposed
detectors, while the black circuit parts are the two JCQs. Here, two
rf-SETs, coupled capacitively (with capacitance $C_c$) to the CBPs,
detect~\cite{SET} the charge states of the qubits. The results of
these $X$-measurements are transferred to the currents (i.e., $I_1$
and $I_2$) biasing the dc-SQUIDs (located at the center of the
figure), which are used to detect~\cite{current} circulating
currents (i.e., perform the $Z$-measurements) along the
inductively-coupled (with mutual inductance $M$) SQUID-loops. Each
one of the voltmeters, $K_i$'s and $S_i$'s ($i=1,2$), detects if the
nearest-neighbor Josephson junction collapses to its normal states.}
\end{figure}
The $X_k$-measurement is achieved, e.g., by a rf-SET
(radio-frequency single-electron transistor)~\cite{SET} coupled
capacitively to the $k$th CPB. Suppose that the applied rf-SET is
sufficiently sensitive to {\it nondestructively} distinguish two
charge states $|0\rangle$ and $|1\rangle$ of the coupled box and the
measured result is then transferred to a current $I$ (or $-I$) if
the measured state is $|0\rangle$ (or $|0\rangle$). Next, the
induced current $I_{k}$ biases the $j$th (but with $j\neq k=1,2$)
dc-SQUID (coupled inductively to the $j$th qubit) for performing the
$Z_k$-measurement: detecting the circulating currents ($|+\rangle$
corresponds to the clockwise current $I^s=I'$, and $|-\rangle$ to
the anticlockwise current $I^s=-I'$) along the $k$th SQUID-loop.

The desirable joint measurements are performed by detecting if the
junctions nearest to the voltmeters, i.e., $K_1, K_2, S_1, S_2$,
collapse to their normal states (this occurs when currents exceed
their critical values~\cite{current}). Suppose that the critical
current $\tilde{I}_c$ of each junction in the two colored dc-SQUIDs
[located at the center of Fig.~2(b)] is set as $|I-I'|<\tilde{I}_c<
|I+I'|$. Let us now focus on the colored dc-SQUIDs. When the two
bias currents $I_1$ and $I_2$ (applied to the two colored SQUIDs)
flow in the same direction (up/down), i.e., $X_1X_2=1$, then
noncontextuality in HVTs predicts that the circulating currents,
$I_1^s$ and $I_2^s$, in the two qubit-loops must be the same
(clockwise/anticlockwise), which implies that $S_i$ and $K_i$
($i=1,2$) should first simultaneously collapse to their normal
states from the superconducting ones. Inversely, the contextuality
of QM predicts that $I_1^s$ and $I_2^s$ must be opposite, and thus
$S_i$ and $K_j$ (with the crucial difference that now $i\neq j=1,2$)
will first simultaneously collapse to the normal states.


This work was supported partly by the NSA, LPS, ARO, NSF grant No.
EIA-0130383; the NSFC grants No. 10874142, No. 60725416 and No.
10625416.

\end{document}